\documentclass{article}
\usepackage{latexsym,amsmath,amsthm,amssymb,amsbsy,graphicx}
\usepackage[cp1251]{inputenc}
\usepackage[T2A]{fontenc}
\usepackage[english]{babel}
\begin{document}

\begin{center}
{ \Large{\bf{Haag's Theorem in Noncommutative Quantum Field Theory
}}}
\end{center}

\vspace{5mm}

\begin{center}
{\large \bf K. V.Antipin$^a$,  M.N.  Mnatsakanova$^b$ and Yu.~S.
Vernov$^c$}

\vspace{5mm}

{\it $^a$Faculty of Physics, Moscow State University,  Moscow, Russia \\
$^b$Skobeltsyn Institute of Nuclear Physics, Moscow State
University, Moscow, Russia \\
$^c$Institute for Nuclear Research, RAS, Moscow, Russia }
\end{center}

\vspace{5mm}

\begin{abstract}
Haag's theorem was extended to noncommutative quantum field theory
in a general case when time does not commute with spatial
variables. It was proven that if S-matrix is equal to unity in one
of two theories related by unitary transformation, then the
corresponding one in another theory is equal to unity as well. In
fact, this result is valid in any $S \, O \, (1,1)$  invariant
quantum field theory, of which an important example is
noncommutative quantum field theory.
\end{abstract}

Pacs{ 11.10.Cd, 11.10.Nx. }

\section{Introduction}\label{int1}
In the present paper we consider Haag's theorem - one of the most
important results of axiomatic approach in quantum field theory
\cite{StrWi,BLT}. In the usual Hamiltonian formalism it is assumed
that asymptotic fields are related with interacting fields by
unitary transformation. Haag's theorem shows that in accordance
with Lorentz invariance of the theory the interacting fields are
also trivial which means that corresponding S-matrix is equal to
unity. So the usual interaction representation can not exist in
the Lorentz invariant theory. Let us recall that Haag's theorem
considers two theories, in which quantum field operators at equal
time as well as corresponding vacuum states $\Psi_0^i$ are related
by the unitary operator:
\begin{gather}
\begin{split}
&\varphi^2_f(t)=V\varphi^1_f(t)V^{-1}\\
&\Psi_0^2=CV\Psi_0^1,\quad C\in \mathbb C, \quad |C|=1
\end{split}
\end{gather}
In accordance with the Haag's theorem four first Wightman
functions coincide in two theories in usual Lorentz invariant case
\cite{StrWi,BLT}. Let us recall that by definition Wightman
functions are:
\begin{equation}
W = W\left(x_1,\ldots,x_n\right) =
\left(\Psi_0,\varphi\,(x_1)\ldots\varphi\,(x_n)\Psi_0\right)
\end{equation}
The main consequence of the Haag's theorem is that from triviality
of one of the fields in question it follows that other one is
trivial too, which implies that corresponding S-matrix is equal to
unity. It is known that in axiomatic quantum field theory (QFT)
there is no field operator defined in a point. Only the smoothed
operators written symbolically as
$$
\varphi_f = \int \varphi\,(t,x)f(x)\,d^3x\,dt,
$$
where $f\,(x)$ are test functions, can be rigorously defined.

In the formulation of Haag's theorem it is assumed that the formal operators
$\varphi\,(t,  x)$ can be smeared only on the spatial variables.

In $S \, O \, (1, k)$ invariant theory, where $k\in\mathbb N$, it
was proved that in two theories related by a unitary
transformation the first $k + 1$ Whigtman functions coincide
\cite{triest}. Thus in $SO(1, 1)$ invariant theory, the most
important example of which is noncommutative quantum field theory,
only the first two Wightman functions coincide.

\section{Noncommutative Quantum Field Theory}\label{sec2}
Noncommutative quantum field theory (NC QFT) being one of the
generalizations of standard QFT has been intensively developed
during the past years (for reviews, see \cite{DN, Sz}). The
present development in this direction is connected with the
construction of noncommutative geometry \cite{Connes} and new
physical arguments in favour of such a generalization of QFT
\cite{DFR}. Essential interest in NC QFT is also due the fact that
in some cases it is a low-energy limit of string theory
\cite{SeWi}.

The simplest and at the same time most studied version of
noncommutative field theory is based on the following
Heisenberg-like commutation relations between coordinates:
\begin{equation}
\label{com}
\left[\hat x^{\mu},\,\hat x^{\nu}\right]=i\theta^{\mu\nu},
\end{equation}
where $\theta^{\mu\nu}$ is a constant antisymmetric matrix.

It is very important that NC QFT can be also formulated in
commutative space, if we replace the usual product of quantum
field operators (strictly speaking, of the corresponding test
functions) by the $\star$- (Moyal-type) product \cite{DN, Sz}.

Let us remind that the $\star$-product is defined as
\begin{equation}
\label{ser} \varphi(x)\star\varphi(y) =
exp\left(\frac{i}2\theta^{\mu\nu}\frac{\partial}{\partial
x^\mu}\frac{\partial}{\partial y^\nu}\right)\varphi(x)\varphi(y)
\equiv
\sum_{n=0}^\infty\frac1{n!}\left(\frac{i}2\theta^{\mu\nu}\frac{\partial}{\partial
x^\mu}\frac{\partial}{\partial y^\nu}\right)^n\varphi(x)\varphi(y)
\end{equation}
Evidently the series in equation (\ref{ser}) has to be convergent.

It was shown \cite{testf} that this series is convergent if the
corresponding test functions belong to one of the Gelfand-Shilov
spaces $S^\beta$ with $\beta < \frac1{2}$.  The similar result was
obtained also in \cite{Sol07}. Moreover,
$\varphi(x)\star\varphi(y)$ belongs to the same space $S^{\beta}$
as $\varphi(x)$ and $\varphi(y)$  \cite{testf}.

Noncommutative theories defined by Heisenberg-like commutation
relations~(\ref{com}) can be divided into two classes.

The first of them is the case of only space-space noncommutativity, that is
$\theta^{0i} = 0$, time commutes with spatial coordinates.

It is known that this case is free from the problems with
causality and unitarity  \cite{nc1,nc2,nc3,nc4,nc5} and in this
case the main axiomatic results: CPT and spin-statistics theorems,
Haag's theorem remain valid  \cite{AGM} - \cite{CMTV}.

Let us remind that if time commutes with spatial coordinates, then
there exists one spatial coordinate, say $x_3$, which commutes
with all others. Simple calculations show that this result is
valid also in a space with arbitrary even number of dimensions.
Thus in space-space NC QFT we have two commuting coordinates and
two noncommuting coordinates. For simplicity we consider
four-dimensional case.

In the second case all coordinates, including time, are noncommuting.

Let us proceed to the LCC in noncommutative space-space QFT.

First let us recall this condition in commutative case. In the operator form
this condition is
\begin{equation}\label{loccom}
\left[\varphi_{f_1},\varphi_{f_2}\right]=0,\quad if\quad O_1\sim O_2,
\end{equation}
where $O_1 = \mbox{supp} \,  f_1, O_2 = \mbox{supp} \,  f_2$. The
condition $O_1 \sim O_2$ means that $(x-y)^2 < 0\, \forall x \in
O_1$ and $y \in O_2$.

In the noncommutative case we have the similar condition with
respect to commutative coordinates, thus now $O_1\sim O_2$ means
that
\begin{eqnarray}
(x_0-y_0)^2-(x_3-y_3)^2<0.
\end{eqnarray}

\section{Haag's theorem in space-space NC QFT}\label{sec3}

Let us recall that in NC QFT Lorentz invariance is broken up to $S
O \, (1,1) \otimes S O \, (2)$ symmetry \cite{AGM}. As was already
mentioned, in two $S O \, (1, 1)$ invariant theories related by a
unitary transformation, only two-point Wightman functions
coincide.

In what follows we prove that if one of considered theories is
trivial, that is the corresponding S-matrix is equal to unity,
then another is trivial too. To prove this in $S \, O (1, 1)$
invariant theory it is sufficient that the spectral condition,
which implies non existence of tachyons, is valid only in respect
with commutative coordinates that is
\begin{equation}\label{spec}
P_i^0 \ge |P_i^3|
\end{equation}
for arbitrary state. \\
Also it is sufficient that translation invariance is valid only in
respect with the commutative coordinates.

Let us point out that in case the spectral condition (\ref{spec})
and $S \, O (1, 1)$ symmetry are fulfilled,  Wightman functions
are analytical functions in two-dimensional analog of extended
tubes \cite{CMTV}.

The equality of two-point Wightman functions in two theories leads
to the following conclusion: if local commutativity condition in
respect with commutative coordinates is fulfilled and the current
in one of the theories is equal to zero, then another current is
zero as well.

Indeed as $W_1(x^1, x^2)=W_2(x^1, x^2)$, then also
\begin{equation}
\left(\Psi^1_0, j^1_{\bar f}j^1_f\Psi^1_0\right) = \left(\Psi^2_0,
j^2_{\bar f}j^2_f\Psi^2_0\right)
\end{equation}
where
\begin{equation*}
j^i_f=\left(\square+m^2\right)\varphi^i_f.
\end{equation*}
If, for example, $j^1_f=0$, then in the space with positive metric
\begin{equation}
j^2_f\Psi^2_0=0
\end{equation}
From the latter formula, corresponding analytical properties of
Wightman functions, and local commutativity
condition~(\ref{loccom}) it follows \cite{BLT} that
\begin{equation}
j^2_f\equiv0
\end{equation}
Our statement is proved.

\section{Haag's Theorem in General Case}\label{sec4}
Now let us consider the general case, when all coordinates,
including time, do not commute. Let us stress that NC QFT in
general case is $S O \, (1, 1)$ invariant as well as in the case
of space-space noncommutativity \cite{AGM}.

In fact our derivation is valid in any theory with $S O \,(1, 1)$
symmetry, if corresponding test functions belong to one of the
Gelfand-Shilov space $S^{\beta}$ with $\beta < \frac1{2}$.

Wightman functions in noncommutative theory take the following form with the Moyal-type product:
\begin{equation}\label{ncwight}
W_\star\,(x_1,\ldots,x_n)=\left(\Psi_0,\phi(x_1)\star\cdots\star\phi(x_n)\Psi_0\right)
\end{equation}
Corresponding generalized function  on  Gelfand-
Shilov space $S^{\beta}$ acts as
\begin{equation}\label{star}
\left(\Psi_0,\phi_{f_1}\star\cdots\star\phi_{f_n}\Psi_0\right) =
\int\, W\,(x_1,\ldots,x_n)f_1(x_1)\star\cdots\star
f_n(x_n)dx_1\ldots dx_n,
\end{equation}
where $W(x_1,\ldots,x_n)=\left(\Psi_0,\phi(x_1)\cdots\phi(x_n)\Psi_0\right)$.

According to~ \cite{testf} expression $f_1(x_1)\star\cdots\star
f_n(x_n)$ is well-defined for $S^{\beta}$ with $\beta < \frac1{2}$
and belongs to the same space $S^{\beta}$ as $f_i(x_i)$.

Let us recall that the series for the Moyal-type product has the
following form:
\begin{gather}
\begin{split}
\label{ncser} f_{\star}(x,y)\equiv
exp\left(\frac{i}2\theta^{\mu\nu}\frac{\partial}{\partial x^\mu}
\frac{\partial}{\partial y^\nu}\right)f(x)f(y)& =
\sum_{n=0}^\infty\frac1{n!}
\left(\frac{i}2\theta^{\mu\nu}\frac{\partial}{\partial
x^\mu}\frac{\partial}{\partial y^\nu}\right)^n\, f(x)\,f(y).
\end{split}
\end{gather}
Here the expression is written only  for two points  $x$ and $y$
instead of $x_1,\ldots, x_n$ for simplicity.

In accordance with eq. (\ref{star}) Wightman functions are
generalized functions corresponding to test functions, which
belong to $S^{\beta}$.

Let us consider the set of functions satisfying the condition
\begin{equation}\label{deriv}
\frac{d^n}{dx^n}f_k(x)=0,\quad \mbox{if} \quad n > k.
\end{equation}
Evidently these functions belong to  $S^{\beta}$. In accordance
with latter condition the expansion (\ref{ncser}) contains a
finite number of derivatives. Thus corresponding  Wightman
functions $W_k$ are tempered distributions.

Now we can take advantage of using $W_k$: at any finite number $k$
we have a local theory, that is a theory with finite maximal speed
of interaction spread. Thus we can assume that just as in the case
of space-space NC QFT quantum field operators satisfy local
commutativity condition in respect with variables satisfying $S O
\,(1, 1)$ symmetry, that gives us possibility to use the same
derivation of Haag's theorem as in the case of space-space NC QFT.
Let us point out that maximal speed of interaction spread depends
on $k$ and goes to infinity if $k\to\infty$, as NC QFT is a theory
with infinite speed of interaction spread.

Thus we have proved Haag's theorem for the special choice of
Wightman functions. Then we can take the limit $k\to\infty$. Let
us take into account that functions $f_k\,(x_1,\ldots,x_n)$,
satisfying eq. (\ref{deriv}), owing to convergence of the set
(\ref{ncser}), go to the limit $f(x_1)\star\cdots\star f(x_n)$. So
the sequence of functionals $W_k$ converges weakly to some
functional $W$. It is known that Gelfand-Shilov space $S^\beta$ is
complete with respect to weak convergence (see~ \cite{GSh},
p.~91).  Haag's theorem is valid for Wightman functions,
coinciding with $W_k$, therefore it is valid for their limit value
$W$ corresponding to $k = \infty$. So Haag's theorem has been
extended to general case of time-space noncommutativity.

\section{Conclusions}\label{sec5}
We see that Haag's theorem is valid in any theory with $S \, O(1,
1)$ symmetry, if corresponding test functions belong to one of the
Gelfand-Shilov spaces $S^{\beta}$ with $\beta < \frac1{2}$. The
important physical example of such a theory is NC QFT.

\end{document}